\documentstyle[prb,preprint,aps]{revtex}

\begin{document}
\draft

\title{Coupling time scales for simulation of structure transformation:
an attempt to combine molecular dynamics and phase-field theory }
\author{Zhi-Rong Liu and Huajian Gao}
\address{ Max Planck Institute for Metals Research, Heisenbergstrasse 3, 
D-70569 Stuttgart, Germany}

\maketitle

\begin{abstract}
A multiscale scheme combining molecular dynamics (MD) and
microscopic phase-field theory is proposed to study the structural
phase transformations in solids with inhomogeneous strain field. The approach
calculates strain response based on MD and atomic diffusion based on the 
phase field theory. Simulations with the new
technique are conducted in two examples. The first involves 
interface roughening in a Co/Cu thin film, where interfacial
undulations due to lattice mismatch is
demonstrated. The second example is a study of spinodal decomposition in
AgCo/Pt/MgO(001) thin film, where we show that Co atoms are
attracted by dislocations in the Pt/MgO interface, 
producing an interesting nanostructure.

\end{abstract}


\vspace{2mm}



Phase-field theory\cite{1} and molecular dynamics\cite{2} (MD) are
both important theoretical tools to investigate the phase
transformation in solids. In particular, phase-field theory has been frequently
used to simulate the structure evolution of micro/mesoscopic
morphology during phase transformations. A key factor to this
capability lies in the flexibility of time scale used in
the theory: one can always choose to trace the evolution
of field variables whose dynamic evolution is ``slow" compared
to the remaining microscopic degrees of freedom. To incorporate
the effect of elastic strains that are responsible for the
formation of various mesoscale morphological patterns, continuum
elasticity theory is usually used to calculate the strain energy
induced by the lattice mismatch.\cite{1,3} However, a priori
assumption on the transformation outcome has to be made because
the elastic constants of the outcome phases are needed in the
calculation. This may seriously reduce the predictive ability
of the phase-field theory (especially the microscopic phase
field model). In addition, the general formula of strain energy
in various situations is often difficult to obtain. The second
shortcoming of the phase-field theory comes from its assumption
of coherent microstructure. It is not suitable for describing
dislocations and their corresponding strain effect at an
incoherent interface between two different phases. On the
other hand, the method of molecular dynamics directly simulates
the motion of moleculars by solving Newton's equations. MD is
highly effective in relaxing the local atomic distances and thus
evaluating the stain energy. MD calculations are able to predict
equilibrium and nonequilibrium properties of condensed systems.
Despite of these advantages, MD has not been used to model
dynamical morphological evolution in phase transformations. The
main problem is that the characteristic time scale of MD is too
short ($\sim$1 ns with time step $\sim$1 fs) compared with that
of most phase transformation phenomena.

In this paper, an approach is developed to combine the microscopic phase-field theory
and molecular dynamics to overcome the above difficulties. Two simple
systems are investigated to illustrate how the new method works.

In MD, the variables to simulate are the atomic positions, \{${\bf r}_i$\},
which evolve according to Newton's equation:\cite{2}
\begin{equation}
m_i\frac{d^2{\bf r}_i}{dt^2}=-\nabla_iV(\{{\bf r}_i\}),
\end{equation}
where $V(\{{\bf r}_i\})$ is the empirical potential. For the
microscopic phase-field theory, the variables considered are the
site occupation probabilities, \{$n_i$\}. The value of $n_i$
represents the probability to find a solute atom at the $i$-th
lattice site. The evolution equation is given as\cite{1}
\begin{equation}
\frac{dn_i}{dt}=\sum_{j}L_{ij}\mu_{j}=\sum_{j}L_{ij}\frac{\partial F(\{n_i\})}{\partial
n_j},
\end{equation}
where $\mu_j$ is the chemical potential at the $j$-th site and
$L_{ij}$ is the kinetic coefficient proportional to the inverse average time
of elementary diffusional jumps from site $i$ to $j$, and $F$ is the
free energy function.

In our method, both the lattice position and the occupation
probability, \{${\bf r}_i,n_i$\}, are considered as variables.
Here, ${\bf r}_i$ are used to describe local atomic motion such as
strain response or dislocation, and $n_i$ are used to describe
atomic diffusion between different lattice sites. With mean-field
approximation, the empirical potential and free energy can be
expressed as functions of both atomic position and
composition, i.e., $V(\{{\bf r}_i,n_i\})$
and $F(\{{\bf r}_i,n_i\})$. For example, when the pair
potential is considered, $V(\{{\bf r}_i,n_i\})$ in the
current approach can be written as:
\begin{equation}
V(\{{\bf r}_i,n_i\})=\frac{1}{2}\sum_{i,j} \left[
n_in_jV_{AA}({\bf r}_i-{\bf r}_j)+2n_i(1-n_j)V_{AB}({\bf r}_i-{\bf
r}_j) +(1-n_i)(1-n_j)V_{BB}({\bf r}_i-{\bf r}_j) \right].
\end{equation}
The evolution of atomic position \{${\bf r}_i$\} is conducted according to Eq.
(1) with the new potential $V(\{{\bf r}_i,n_i\})$. When the elastic
equilibrium is reached, \{${\bf r}_i$\} satisfy the following equations:
\begin{equation}
\frac{\partial V(\{n_i,{\bf r}_i\})}{\partial {\bf r}_i}=0.
\end{equation}
Since the free energy depends on \{${\bf r}_i$\} only via $V(\{{\bf r}_i,n_i\})$,
one has
\begin{equation}
\frac{\partial F(\{n_i,{\bf r}_i\})}{\partial {\bf r}_i}=0.
\end{equation}
Since diffusion is much slower than elastic response, we can assume that
the system is in elastic equilibrium during the diffusion process.
The diffusion equations are:
\begin{eqnarray}
\nonumber \frac{dn_i}{dt}=\sum_{j}L_{ij}\mu_{j}
& = & \sum_{j}L_{ij}\frac{d F(\{{\bf r}_i,n_i\})}{d n_j} \\
\nonumber
& = & \sum_{j}L_{ij}\left[\frac{\partial F(\{{\bf r}_i,n_i\})}{\partial n_j}
+\sum_{k}\frac{\partial F(\{{\bf r}_i,n_i\})}{\partial {\bf r}_k}
\cdot \frac{d{\bf r}_k}{d n_j} \right]\\
& = & \sum_{j}L_{ij} \frac{\partial F(\{{\bf r}_i,n_i\})}{\partial n_j},
\end{eqnarray}
where ${d F(\{{\bf r}_i,n_i\})}/{d n_j}$ means that \{${\bf
r}_i$\} is function of $n_j$ and the differential with respect to
$n_j$ should also operat on ${\bf r}_i$. The elastic
equilibrium conditions Eq. (5) is used to obtain the final result
of Eq. (6). In the numerical scheme, $m$ MD steps with time
interval $\delta t$ are conducted to reach the elastic equilibrium
state before every diffusion step with time interval $dt$. It is a
multiscale scheme since $m\delta t\ll dt$.

In order to illustrate how the method works in practice, we first
consider interface roughening in a heteroepitaxial thin film system.

In uniform heteroepitaxial thin films, there usually exists large stress
arising from the lattice mismatch between thin film and
substrate. To reduce the strain energy, surface roughening by mass
diffusion occurs during film growth or annealing.\cite{4}
The length scale of surface roughening is about a few hundred
nanometers,\cite{4} which is too large for the current atomic simulation.
We alternatively simulate a similar process, the interface roughening,
to demonstrate the effect of strain on the mass transport.

The system studied here is Cu-Co system where the lattice mismatch
is about 2\%. Cu and Co is immiscible in the bulk with a mixing
energy of +13 kJ/g atom.\cite{4a} Atomic interaction among Cu and Co atoms is
described with the Tight-binding second-moment-approximation
(TB-SMA) potential.\cite{5} The potential parameters for the pure
species are available in the literature,\cite{5} while the cross
parameters (Cu-Co) are evaluated by fitting the mixing energy and
the lattice length of the disordered phase within the mean-field
approximation. We consider a coherent Co film grown on Cu substrate.
The as-grown film (initial state) has flat surface and interface
[Fig. 1(a)]. The evolution of phase morphology is simulated
by the above method, with result shown in Fig. 1(b). It 
reveals interfacial undulations with varying amplitude and
wavelength. The undulation of interfacial increases the
interfacial energy while decreases the strain energy. The total
energy decreases in such process. In comparison with surface
roughening,\cite{4,6} no groove or cusp is found in the current
simulation of interface roughening.


As the next example, we consider the spinodal decomposition of a thin film on a periodically
strained substrate.

Adsorbed layers epitaxially grown on
single-crystal surfaces are generally strained due to the lattice
mismatch. In the weakly incommensurate situation such as Ag layers
grown on Pt(111) surface,\cite{7} highly ordered periodic
dislocation network is formed to relieve the mismatch strain,
hence invalidating the assumption of coherent microstructure 
usually adopted in the phase field theory. Such dislocation pattern can be used to fabricate
interesting metal island arrays in two-dimension.\cite{8} If the
second alloy film is grown on this strained film and the second
film is unstable to spionodal decomposition, it may be expected
that the dislocation network will act as a preferred length scale
for the spinodal decomposition. To investigate this issue, we
simulate the spinodal decomposition in a AgCo/Pt/MgO(001) thin film system.

For Pt films deposited onto a (001) MgO substrate, a dislocation
network with periodicity 4.05 nm is predicted.\cite{9} We consider
the growth of AgCo on the strained Pt/MgO thin film and simulate
the spinodal decomposition of the AgCo phase. The potential
between Ag-Co-Pt is adopted as TB-SMA potential.\cite{5} For simplicity,
the interaction between Pt and MgO is approximated as the Lenard-Jones
potential by fitting the adsorption energy and atomic
distance.\cite{10} MgO atoms are kept fixed (with crystal lattice
of 4.21 \AA) in simulation. For the initial state,
Pt atoms are arranged in their ideal lattice sites (with a lattice
parameter of 3.924 \AA), and a dislocation network between Pt and
MgO naturally forms in the simulation. The initial AgCo phase
is assumed to be coherent with Pt and is disordered with an average
composition $n_i=0.5$. A simulation result of the thin film with 3
monolayer (ML) of Pt and 10 ML of AgCo is shown in Fig. 2. The
atomic configuration with dislocation network and the
corresponding strain effect are demonstrated in Fig. 2(a) and 2(b)
for the initial stage before atomic diffusion occurs. It can be
seen that large strain is concentrated near the dislocation cores.
The final result of phase transformation is depicted as Fig. 2(c).
It shows that Co atoms segregate in strained regions because they are
smaller in size. Within a few layers near the AgCo/Pt interface,
Ag atoms gather in less strained regions and produce an
interesting nanostructure. The surface layers are occupied by Ag
atoms since the surface energy of Ag (1.30 J/m$^2$) is smaller
than that of Co atoms (2.71 J/m$^2$).\cite{11}
It is interesting that some Pt atoms
above the dislocation are ``pulled up" by Co atoms in Fig. 2(c), which reflects the
important interaction between Pt substrate and AgCo film. 
Such effect can not be described within the coherency approximation.


The method presented here is on the microscopic level. Similar to other atomic scale simulations, one of its disadvantage is that the system size simulated
is relatively small compared to mesoscale morphology. In contrast, continuum field theories have more freedom in changing the length scale of simulation.
Such theories can be expected to couple with computational mechanics methods
such as the Finite Element method by a similar coupling approach as described
here. This is not pursued in this paper.


This work was supported by a Max Planck Post-Doc Fellowship for
ZL and by US National Science Foundation through Grant CMS-0085569.
We thank Prof. Long-Qing Chen at the Pennsylvania State University
for helpful discussions. The code of MD simulation in this work
comes from the program GONZO of Stanford Multiscale Simulation
Laboratory (MSL).

\begin{figure}[tbp]
\caption{Interface roughening for a Co film grown on Cu substrate.
The Cu phase is colored in red while the Co phase blue.
The axes have length unit of \AA. The system is composed of
$200\times 60 \times 1$ atoms. Periodic boundary conditions
are adopted in the $x$- and $z$-directions while free boundary
conditions in the $y$- direction. (a) The as-grown film
(initial state of simulation). (b) The simulation result
of interface roughening.}
\end{figure}

\begin{figure}[tbp]
\caption{Dislocation, strain effect and spinodal decomposition in
a AgCo/Pt/MgO(001) thin film. (a) Atomic configuration of the
as-grown film before the spinodal decomposition of Ag-Co occurs.
The blue hollow circle ($\circ$) and cross ($+$) represent O and
Mg atoms, respectively. The green stars ($\ast$) are Pt atoms,
and the red filled circles ($\bullet$) are disordered AgCo
atoms with occupation probability $n_i=0.5$. (b) The atomic
displacement of the as-grown film in (a). Strain field is
caused by dislocations between Pt and MgO. (c) The resulting
configuration after phase transformation. The occupation
probability of Ag-Co phase is visualized by the color scheme shown
in the colorbar.
}
\end{figure}


\vspace{2mm}

\begin{references}

\bibitem{1} A. G. Khachaturyan, {\it Theory of Structural
Transformations in Solids} (Wiley, New York, 1983).

\bibitem{2} M. P. Allen and D. J. Tildesley, {\it Computer Simulation of
Liquids,} (Clarendon Press, Oxford, 1996).

\bibitem{3} S. Y. Hu and L. Q. Chen, Acta Mater. {\bf 49}, 1879 (2001).

\bibitem{4} H. Gao and W. D. Nix, Annu. Rev. Mater. Sci. {\bf 29}, 173 (1999).

\bibitem{4a} C. Gente, M. Oehring, and R. Bormann, Phys. Rev. B {\bf 48}, 13244 (1993).

\bibitem{5} F. Cleri and V. Rosato, Phys. Rev. B {\bf 48}, 22 (1993).

\bibitem{6} W. H. Yang and D. J. Srolovitz, Phys. Rev. Lett. {\bf 71}, 1593 (1993).

\bibitem{7} H. Brune, H. Roder, C. Boragno, and K. Kern, Phys. Rev. B
{\bf 49}, 2997 (1994).

\bibitem{8} H. Brune, M. Giovannini, K. Bromann, and K. Kern, Nature
{\bf 394}, 451 (1998).

\bibitem{9} P. C. Mcintyre, C. J. Maggiore, and M. Nastasi, Acta Mater.
{\bf 45}, 869 (1997).

\bibitem{10} A. Bogicevic and D. R. Jennison, Surf. Sci. {\bf 437}, L741 (1999).

\bibitem{11} L. Z. Mezey and J. Giber, Jpn. J. Appl. Phys. {\bf 21}, 1569 (1982).

\end{references}
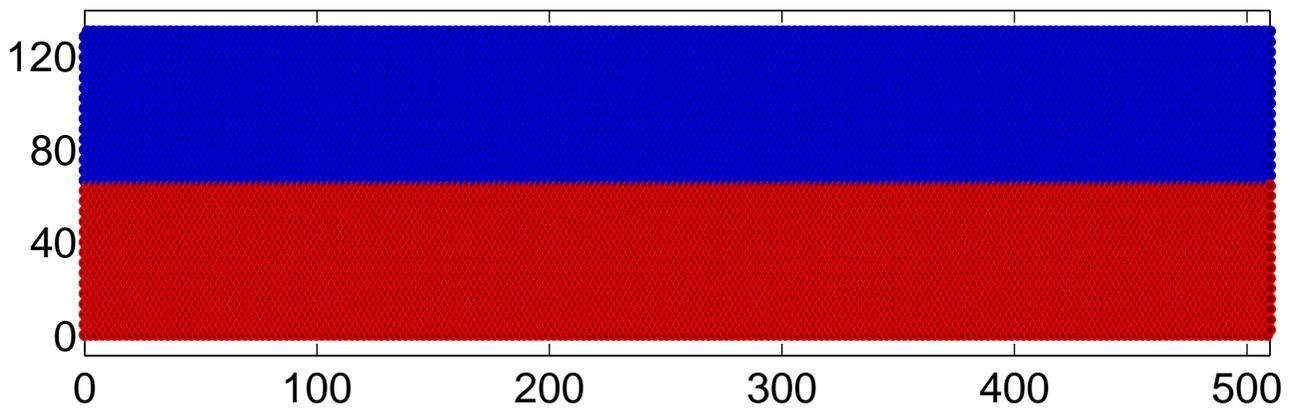
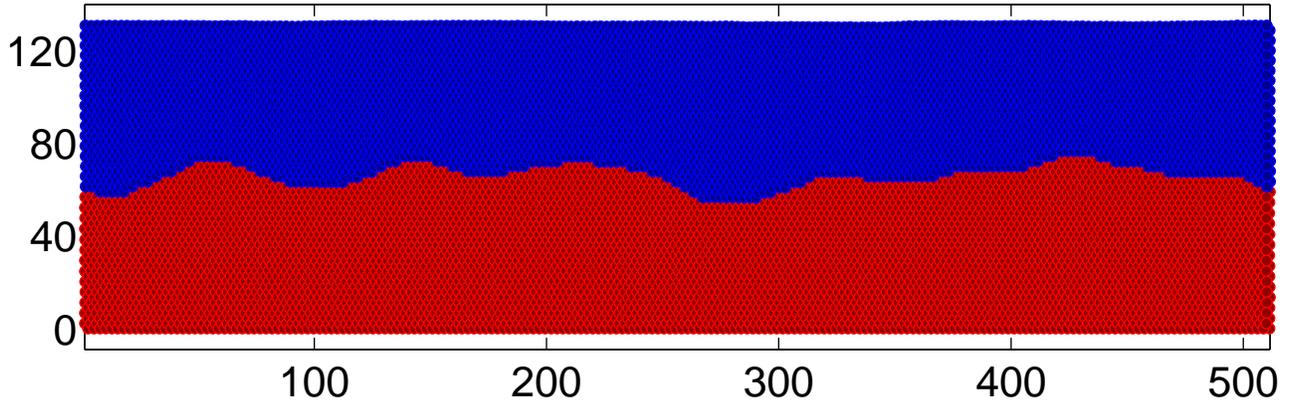
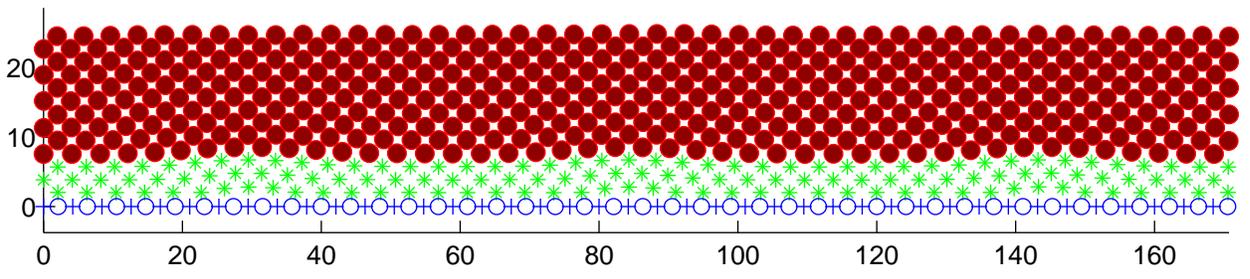
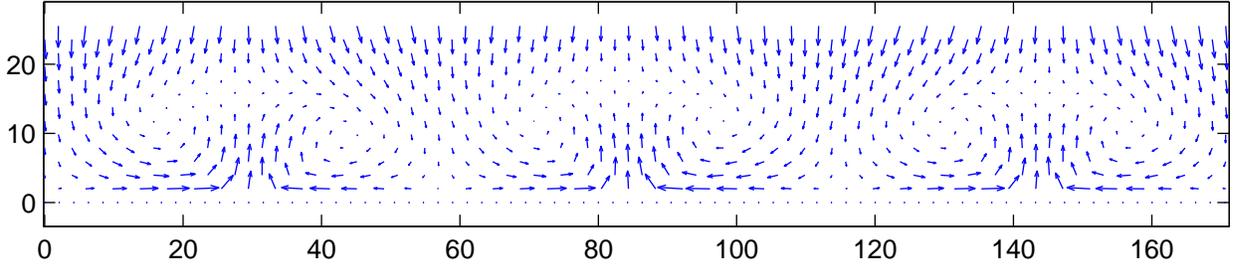
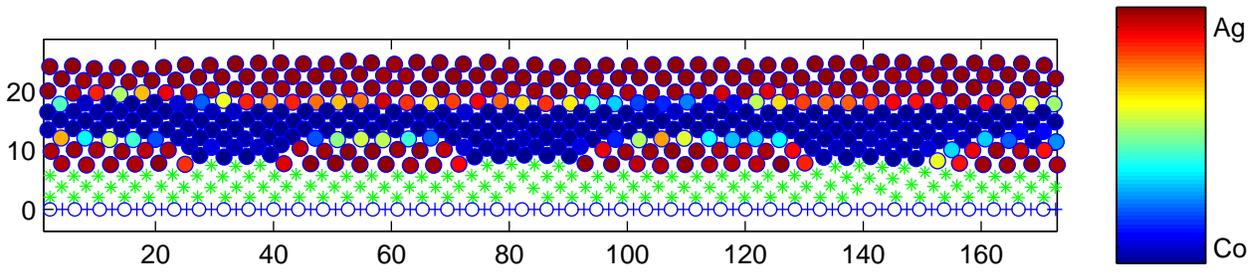
\end{document}